# The noise fluxes produced by the degree of first-order temporal coherence in a single mode class-A laser amplifier

S. Kiashemshaki and J. Jahanpanah

Physics Faculty, Kharazmi University, 49 Mofateh Ave, 15719-14911, Tehran, Iran

E-mail: jahanpanah@khu.ac.ir

**Abstract**

The noise feature of a single mode class-A laser is investigated in the presence (amplifier) and absence (free-running) of an input signal. The Maxwell-Bloch equations of motion have been solved after adding the cavity Langevin force to calculate fluctuations that imposed to the atomic population inversion and the amplitude and phase of cavity electric field. The correlation function of these fluctuations is then used to derive the spontaneous emission, amplitude, and phase noise fluxes in the below and above-threshold states. The bandwidth of noise fluxes is not only adjusted by the amplitude and frequency detuning of input signal, but also by the laser pumping and cavity damping rates. On the other hand, the degree of first-order temporal coherence (DFOTC) is turned out as the correlation function of the amplitude fluctuation so that its Fourier transform led to the amplitude noise flux. The coherence time plays a dual role in order that it is equal to the damping rate invers of DFOTC and at the same time has an uncertainty relation with the bandwidth of amplitude noise flux. Finally, the flux conservation requires a balance between the input pumping noise flux and the output amplitude and spontaneous emission noise fluxes.

## 1. Introduction

The amplification (gain) mechanism of laser amplifiers in the both below and above-threshold states has so far been reported by many literatures [1-4]. The above-threshold state is distinguished by the two different regimes of single frequency (injection-locked) and multiple frequency [5-7]. The multiple frequency regime consists of many signal and image satellite lines, which are produced by the four-wave mixing interaction between the amplified input signal and the strong cavity electric field [8, 9]. By increasing the intensity or reducing the frequency detuning of input signal just at the common border of two regimes, the injection-locking phenomenon suddenly happens so that the frequencies of slave cavity electric field, signal, and image satellite lines are simultaneously captured by that of master input signal [8, 10]. Therefore, the below-threshold state and the injection-locked regime of laser amplifier exactly mimic the above-threshold state of a free-running laser due to a single-component cavity electric field which oscillates at the input signal rather cavity resonance frequency [2, 11].

The noise aspect of laser amplifiers has also been studied extensively by solving the laser amplifier equations of motion in the presence of two Langevin (fluctuating) forces of the atomic population inversion and the atomic dipole moment [12-14]. The third Langevin force is cavity one which has always been ignored due to the negligible number of thermal photons inside the laser cavity [12, 13]. The first contradictory proof published in 2012 where the Maxwell-Bloch equations of motion solved in the presence of cavity Langevin force alone [15]. It was demonstrated that the cavity Langevin force is not only a negligible force but also it is able to generate the different frequency spectra of a single-peak Lorentzian and double-peak profiles for the respective amplitude noise fluxes of free-running class-A and -B lasers in complete agreement with the experimental and theoretical results [10, 13, 15, 16].

The aim of present paper is to reveal the importance of cavity Langevin force in more details by calculating the noise fluxes of a class –A laser amplifier in the below-threshold state and injection-locked regime. We consider the input signal as an ideal coherent light with the fixed amplitude and phase (no fluctuation) which injected into a free-running laser cavity for the amplification purpose. In spite of, it is observed that the amplitude and phase of the output amplified signal together with the static atomic population inversion seriously suffer from the fluctuations that imposed by the laser pumping system [17-19]. Therefore, our first priority is to determine these fluctuations and their corresponding noise fluxes by solving the laser amplifier equations of motion in the presence of cavity Langevin force. It is demonstrated that the input noise flux of pumping supplies the output noise fluxes of amplitude and spontaneous emission according to the flux conservation law.

The next priority is to calculate the coherence time of light radiated from the laser amplifier and to recognize the laser and input signal parameters which are involved in the coherence time [20]. We begin with the degree of first-order temporal coherence (DFOTC) which is defined as the correlation function of the amplitude fluctuation of cavity electric field. The main advantage of DFOTC is to relate with the amplitude noise flux by the Fourier transform. It is also turned out that the damping rate of DFOTC has a revers relation with the coherent time [21]. On the other hand, the coherence time demonstrates an uncertainty relation with the common bandwidth of amplitude and spontaneous emission noise fluxes so that they can simultaneously be adjusted by the four key parameters of the amplitude and frequency detuning of the input signal, the laser pumping rate, and the mean damping rate of cavity mirrors.

The noise fluxes of free-running class-A laser are directly extracted from those of class-A laser amplifier by tending the amplitude and frequency detuning of the input signal toward zero. It will be illustrated that the bandwidth of noise fluxes and the coherence time have reversely changed by increasing the laser pumping rate from the below to above-threshold state.

The manuscript is arranged according to the following sections. The motion equations and trial solutions of a class-A laser amplifier together with the zero and first-order solutions are presented in section 2. The output amplitude, spontaneous emission noise fluxes and the input pumping noise flux are first calculated and then demonstrated to satisfy the flux conservation in section 3. The section 4 is allocated to the free-running noise fluxes and their bandwidths which are directly derived from the laser amplifier relations of section 3 after ignoring from the input signal parameters. The DFOTC, coherence time, and their relations with the amplitude noise flux are clarified for the laser amplifier in section 5 and for the free-running laser in section 6. The results are summarized in section 7.

## 2. The motion equations and trial solutions

The general feature of class-A laser amplifiers is described by the two motion equations of Maxwell-Bloch in the forms [13, 15, 22, 23]

$$\dot{\alpha} + (\gamma_C + i\omega_L)\alpha = \frac{g^2}{\gamma_\perp}\alpha D + \Gamma_\alpha + \gamma_2^{1/2}\beta_{in} \qquad (1)$$

and

$$\gamma_\| D = \gamma_\| D_P - \frac{2g^2}{\gamma_\perp}|\alpha|^2 D, \qquad (2)$$

in which the variables $\alpha$ and $D$ are the cavity electric field and atomic population inversion with the respective damping rates $\gamma_C$ and $\gamma_\|$. $\gamma_\perp$ is the damping rate of atomic dipole moment whose equation of motion has adiabatically been eliminated due to the damping condition $\gamma_\perp \gg \gamma_\| \gg \gamma_C$ of class-A lasers. $\gamma_\| D_P$ is the laser pumping rate whose energy is partially converted to the spontaneous emission radiation by the rate $\gamma_\| D$. $\omega_L$ and $g$ are the cavity resonance frequency and coupling constant between the cavity electric field and atomic dipole moment, respectively. The cavity Langevin force is denoted by $\Gamma_\alpha$ with a zero-mean value $<\Gamma_\alpha> = 0$ and the following correlation function [15]

$$<\Gamma_\alpha(\omega)\Gamma_\alpha^*(\omega')> = 2\gamma_C(n_{th}+1)\delta(\omega-\omega') \approx 2\gamma_C \delta(\omega-\omega'), \qquad (3)$$

where the correlation function relation $<\Gamma_\alpha^*(\omega')\Gamma_\alpha(\omega)> = 2\gamma_C n_{th}\delta(\omega-\omega')$ is ignored due to the negligible number of thermal photons ($n_{th} \ll 1$) inside the laser cavity [15, 16]. Finally, the input signal $\beta_{in}$ is injected through the one of cavity mirrors with the amplitude loss rate $\gamma_2^{1/2}$ for the amplification purpose. It is characterized by the two key parameters of amplitude $\beta_S$ and frequency detuning $\omega_d = \omega_S - \omega_L$ in the form

$$\beta_{in} = \beta_S \exp(-i\omega_S t) = \beta_S \exp[-i(\omega_L + \omega_d)t]. \qquad (4)$$

We now define the trial solutions for the single-mode cavity electric field $\alpha(t)$ and the atomic population inversion $D(t)$ as

$$\alpha(t) = [\alpha_S + \delta\alpha_S(t)]\exp[-i(\omega_L + \omega_d)t + i\delta\phi_S(t)] \qquad (5)$$

and

$$D(t) = D_S + \delta D_S(t), \qquad (6)$$

in which $\alpha_S$ and $D_S$ are, respectively, the amplitude of cavity electric field and the statics atomic population inversion which are fluctuated by the real values $\delta\alpha_S(t)$ and $\delta D_S(t)$ due to the cavity

Langevin (fluctuating) force $\Gamma_\alpha$. $\delta\phi_S(t)$ is the third real fluctuating variable associated with the phase of cavity electric field.

By substituting the trial solutions (5) and (6) into the Maxwell-Bloch equations of motion (1) and (2), a cubic equation is turned out for the normalized mean number of photons $|\alpha_S|^2/n_S$ inside the laser cavity correct to the zero-order fluctuation ($\delta\alpha_S \approx \delta D_S \approx \delta\phi_S \approx 0$) as [24]

$$[1+(\omega_d/\gamma_C)^2]\left(\frac{|\alpha_S|^2}{n_S}\right)^3 + \left[2(1-C)+2(\omega_d/\gamma_C)^2 - \frac{\gamma_2|\beta_S|^2}{n_S}\right]\left(\frac{|\alpha_S|^2}{n_S}\right)^2$$
$$+\left[(1-C)^2+(\omega_d/\gamma_C)^2 - 2\frac{\gamma_2|\beta_S|^2}{n_S}\right]\frac{|\alpha_S|^2}{n_S} - \frac{\gamma_2|\beta_S|^2}{n_S} = 0, \tag{7}$$

in which the normalized pumping rate $C = \gamma_\| D_P / \gamma_\| D_0$ has a value less than one in the below-threshold state ($C<1$), equal to one at the threshold state ($C=1$), and larger than one in the above-threshold state ($C>1$). $D_0 = \gamma_\perp \gamma_C/g^2$ is the statics population inversion in the above-threshold state of free-running laser, and $n_S = \gamma_\perp \gamma_\|/2g^2$ is the number of cavity photons in the saturation state at the normalized pumping rate $C=2$ [8, 13]. The normalized population inversion of laser amplifier $D_S/D_0$ is then calculated by substituting $|\alpha_S|^2/n_S$ from the cubic equation (7) into the following relation [8]

$$\frac{D_S}{D_0} = \frac{C}{1+|\alpha_S|^2/n_S}. \tag{8}$$

On the other side, the motion equations for the three fluctuating variables $\delta\alpha_S$, $\delta D_S$, and $\delta\phi_S$ are similarly derived by substituting the trial solutions (5) and (6) into the Maxwell-Bloch equations of motion (1) and (2), but by considering the terms correct to the first-order fluctuation ($\delta\alpha_S^2 \approx \delta D_S^2 \approx \delta\phi_S^2 \approx 0$) as

$$\delta\dot\alpha_S + i|\alpha_S|\delta\dot\phi_S + \gamma_C\left(1-\frac{D_S}{D_0}-i\frac{\omega_d}{\gamma_C}\right)\delta\alpha_S - \frac{g^2}{\gamma_\perp}|\alpha_S|\delta D_S = \Gamma_\alpha \exp(i\omega_s t - i\delta\phi_S) \tag{9}$$

and

$$\gamma_\|\left(1+\frac{|\alpha_S|^2}{n_S}\right)\delta D_S = -4\gamma_C C|\alpha_S|\delta\alpha_S. \tag{10}$$

Now by taking the Fourier transform of equation (9), it is separated into the real and imaginary parts in frequency domain as

$$-i\omega\delta\alpha_S(\omega) + \gamma_C\left(1 - \frac{D_S}{D_0}\right)\delta\alpha_S(\omega) - \frac{g^2}{\gamma_\perp}|\alpha_S|\delta D_S(\omega) = 0.5\left[\Gamma_\alpha(\omega_S + \omega) + \Gamma_\alpha^*(\omega_S + \omega)\right] \quad (11)$$

and

$$\omega|\alpha_S|\delta\phi_S(\omega) - i\omega_d\,\delta\alpha_S(\omega) = 0.5\left[\Gamma_\alpha(\omega_S + \omega) - \Gamma_\alpha^*(\omega_S + \omega)\right]. \quad (12)$$

The three variables $\delta\alpha_S(\omega)$, $\delta\phi_S(\omega)$, and $\delta D_S(\omega)$ will ultimately be rendered from the simultaneous solution of equations (10)-(12) in the forms

$$\delta\alpha_S(\omega) = \frac{\Gamma_\alpha(\omega_S + \omega) + \Gamma_\alpha^*(\omega_S + \omega)}{2(\gamma_C B - i\omega)}, \quad (13)$$

$$\delta\phi_S(\omega) = \frac{(\gamma_C B - i\omega)\left[\Gamma_\alpha(\omega_S + \omega) - \Gamma_\alpha^*(\omega_S + \omega)\right] + i\omega_d\left[\Gamma_\alpha(\omega_S + \omega) + \Gamma_\alpha^*(\omega_S + \omega)\right]}{2\omega|\alpha_S|(\gamma_C B - i\omega)}, \quad (14)$$

and

$$\delta D_S(\omega) = \frac{-2\gamma_C C\,|\alpha_S|\left[\Gamma_\alpha(\omega_S + \omega) + \Gamma_\alpha^*(\omega_S + \omega)\right]}{\gamma_\|(\gamma_C B - i\omega)\left(1 + |\alpha_S|^2/n_S\right)^2}, \quad (15)$$

where

$$B = 1 - \frac{D_S}{D_0} + 2C^{-1}\left(\frac{|\alpha_S|^2}{n_S}\right)\left(\frac{D_S}{D_0}\right)^2, \quad (16)$$

is a dimensionless quantity so that $2B$ will represent the normalized common bandwidth of amplitude, spontaneous emission, and pumping noise fluxes in the following section.

## 3. Balance between the noise fluxes of amplitude $N_{AM}^{LA}(\omega)$, spontaneous emission $N_{SP}^{LA}(\omega)$, and pumping $N_{Pump}^{LA}(\omega)$

Assume that $a(\omega)$ to be an arbitrary fluctuating variable with a white noise origin (Dirac function), then it is required to obey the following correlation function in a complex conjugate form [25]

$$<a^*(\omega)a(\omega')> = 2\pi h(\omega)h^*(\omega')\delta(\omega - \omega'), \quad (17)$$

where $|h(\omega)|^2$ is the dimensionless mean flux per unit angular frequency bandwidth at angular frequency $\omega$. It is evident from the solutions (13)-(15) that the relation (17) is applicable for the three fluctuating variables of cavity electric field amplitude $\delta\alpha_S(\omega)$, phase $\delta\phi_S(\omega)$, and atomic population inversion $\delta D_S(\omega)$ because they are produced by the cavity Langevin force $\Gamma_\alpha$ with a correlation function in the form of Dirac function as defined by (3).

As a result, the amplitude noise flux inside the laser cavity $\left|h_{AM}^{LA}(\omega)\right|^2$ can be calculated by substituting $\delta\alpha_S(\omega)$ from (13) into the correlation function relation (17). The amplitude noise flux of laser amplifier $N_{AM}^{LA}(\omega)$ that emerged from the cavity mirrors of the total damping rate $2\gamma_C = \gamma_1 + \gamma_2$ is thus given by

$$N_{AM}^{LA}(\omega) = 2\gamma_C \left|h_{AM}^{LA}(\omega)\right|^2 = \frac{\gamma_C^2}{2\pi\left(\omega^2 + \gamma_C^2 B^2\right)}, \tag{18}$$

where it has a Lorentzian profile with a normalized bandwidth equal to

$$\frac{(\Delta\omega)_{AM}^{LA}}{\gamma_C} = 2B = 2\left[1 - \frac{D_S}{D_0} + 2C^{-1}\left(\frac{|\alpha_S|^2}{n_S}\right)\left(\frac{D_S}{D_0}\right)^2\right]. \tag{19}$$

It should be noticed that the variables of $|\alpha_S|^2/n_S$ and $D_S/D_0$ are numerically calculated from the cubic equation (7) and the relation (8), respectively. Accordingly, the amplitude noise flux (18) and its normalized bandwidth (19) can be adjusted by choosing the three key parameters of the normalized laser pumping rate $C$ and the normalized rates of input signal flux $\gamma_2|\beta_S|^2/n_s$ and frequency detuning $\omega_d/\gamma_C$. Clearly, the effect of mean damping rate of cavity mirrors $\gamma_C$ as the fourth parameter is hidden in the normalized quantities of amplitude bandwidth $(\Delta\omega)_{AM}^{LA}/\gamma_C$, frequency detuning $\omega_d/\gamma_C$, and Fourier frequency $\omega/\gamma_C$.

The amplitude noise flux $N_{AM}^{LA}(\omega)$ against the normalized Fourier frequency $\omega/\gamma_C$ is plotted in Fig. 1 for the typical normalized input signal flux $\gamma_2|\beta_S|^2/n_s = 0.25$, the different normalized frequency detuning $\omega_d/\gamma_C = 0.25, 0.5, 0.75$, and 1, and the normalized pumping rates (a)- $C = 0.5$ associated with the below-threshold state and (b)- $C = 1.2$ associated with the above-threshold state. The Lorentzian profiles are apparent in the both below and above-threshold states in agreement with the empirical results (please see Fig. 3b of Refs [10] and [19]) and the free-running case (please see Figs. 1 and 2 of Ref [15]). Moreover, the effect of the amplitude and frequency detuning of input signal is clarified in Fig. 2 where the normalized bandwidth of amplitude noise flux $(\Delta\omega)_{AM}^{LA}/\gamma_C$ is plotted versus the normalized frequency detuning $\omega_d/\gamma_C$ for the different normalized input signal flux $\gamma_2|\beta_S|^2/n_s = $ 0, 0.025, 0.25, 0.5, and 0.75, and for the normalized pumping rates (a)- $C=0.5$ and (b)- $C=1.2$. By comparing Figs. 1 and 2, it is evident that the bandwidth (in contrast with the height) of amplitude noise flux is decreased by raising the normalized frequency detuning $\omega_d/\gamma_C$ in the both below-threshold state and injection-locked regime of laser amplifier. The single values 1 and 0.67 which are illustrated in the respective Figs. 2(a) and 2(b) are associated with the amplitude noise bandwidth of the below ($C=0.5$) and above ($C=1.2$) threshold states of free-running laser with the input signal parameters $\gamma_2|\beta_S|^2/n_s=0$ and $\omega_d/\gamma_C=0$. They are calculated according to the corresponding relations (30) and (32) which will be presented in the coming section.

Similarly, the noise fluxes of phase and spontaneous emission are respectively derived by substituting their respective fluctuating variables $\delta\phi_S(\omega)$ and $\delta D_S(\omega)$ from (14) and (15) into the correlation function relation (17) as

$$N_{PH}^{LA}(\omega) = 2\gamma_C|\alpha_S|^2|h_{PH}^{LA}(\omega)|^2 = \frac{\gamma_C^2 B^2 + (\omega+\omega_d)^2}{\omega^2} N_{AM}^{LA}(\omega) \qquad (20)$$

and

$$N_{SP}^{LA}(\omega) = \frac{|h_{SP}^{LA}(\omega)|^2}{2\gamma_C|\alpha_S|^2} = \frac{4C^2}{(1+|\alpha_S|^2/n_S)^4} N_{AM}^{LA}(\omega). \qquad (21)$$

It is evident that the phase noise flux (20) has no special profile, whereas the spontaneous emission noise flux (21) has a Lorentzian profile with the same normalized bandwidth $2B$ of amplitude noise flux $N_{AM}^{LA}(\omega)$ given by (19).

The role of laser pumping is to supply the required energy for the amplification of an input signal according to the following energy conservation relation (please see (2.13) of Ref [8])

$$2\gamma_C|\alpha(t)|^2 - \gamma_2^{1/2}[\alpha^*(t)\beta_{in} + \alpha(t)\beta_{in}^*] + \gamma_\| D(t) = \gamma_\| D'_P(t), \qquad (22)$$

in which $\gamma_\| D'_P(t)$ refers to the input energy rate into laser amplifier by the pumping system which includes the static mean value $\gamma_\| D_P$ together with the fluctuating value $\gamma_\| \delta D_P(t)$. The energy conservation (22) is then transformed to a conservation relation for the fluctuating variables $\delta\alpha_S(t)$, $\delta D_S(t)$, and $\delta D_P(t)$ after substituting the trial solutions (5) and (6) as

$$4\gamma_C |\alpha_S| \delta\alpha_S(t) - 2\gamma_\parallel^{1/2} \beta_S \delta\alpha_S(t) + \gamma_\parallel \delta D_S(t) = \gamma_\parallel \delta D_P(t), \quad (23)$$

so that $\delta D_P(t)$ acts as a fluctuating source for the other two fluctuating variables $\delta\alpha_S(t)$ and $\delta D_S(t)$. The above fluctuation conservation relation is reduced to that of free-running laser given by (39) of Ref [15] after ignoring from the input signal amplitude $\beta_S$.

If one now takes the Fourier transform of (23) and multiplies in its complex conjugate, then a flux conservation relation is derived for the noise fluxes of amplitude $N_{AM}^{LA}(\omega)$, spontaneous emission $N_{SP}^{LA}(\omega)$, and pumping $N_{Pump}^{LA}(\omega)$ by using the correlation function (3) as

$$4S(\beta_S, \omega_d, C) N_{AM}^{LA}(\omega) + N_{SP}^{LA}(\omega) = N_{Pump}^{LA}(\omega), \quad (24)$$

in which

$$S(\beta_S, \omega_d, C) = 1 - \frac{\beta_S/\sqrt{\gamma_C n_S}}{|\alpha_S|/\sqrt{n_S}} + \frac{\beta_S^2/\gamma_C n_S}{4|\alpha_S|^2/n_S} - \left(2 - \frac{\beta_S/\sqrt{\gamma_C n_S}}{|\alpha_S|/\sqrt{n_S}}\right) \frac{D_S/D_0}{1+|\alpha_S|^2/n_S}. \quad (25)$$

It is emphasized that the normalized pumping noise flux will independently be derived from the Fourier transform of fluctuation conservation (23) as

$$N_{Pump}^{LA}(\omega) = \frac{\langle [\gamma_\parallel \delta D_P^*(\omega)][\gamma_\parallel \delta D_P(\omega)] \rangle}{2\gamma_C |\alpha_S|^2} = \left(2 - \frac{\beta_S/\sqrt{\gamma_C n_S}}{|\alpha_S|/\sqrt{n_S}} - 2\frac{D_S/D_0}{1+|\alpha_S|^2/n_S}\right) N_{AM}^{LA}(\omega). \quad (26)$$

Finally, the noise flux conservation (24) can analytically be testified by substituting the noise fluxes of amplitude $N_{AM}^{LA}(\omega)$, spontaneous emission $N_{SP}^{LA}(\omega)$, and pumping noise $N_{Pump}^{LA}(\omega)$ from (18), (21), and (26), respectively.

## 4. The noise fluxes of free-running class-A laser

It is expected that the noise fluxes of free-running laser are directly extracted from those of laser amplifier by applying the input signal amplitude $\beta_S = 0$ and frequency detuning $\omega_d = 0$. Lets first consider the simpler case of the below-threshold state $(C<1)$ so that in the absence of the cavity electric field $\alpha_S = 0$, the atomic population inversion is reduced to $D_S = CD_0$ according to the relation (8). The noise fluxes of amplitude (18), spontaneous emission (21), and pumping (26) are then simplified to

$$N_{AM}^{BFL}(\omega) = \frac{\gamma_C^2}{2\pi[\omega^2 + \gamma_C^2(1-C)^2]}, \quad (27)$$

$$N_{SP}^{BFL}(\omega) = 4C^2 N_{AM}^{BFL}(\omega),  \qquad (28)$$

and

$$N_{Pump}^{BFL}(\omega) = 4(1-C)^2 N_{AM}^{BFL}(\omega),  \qquad (29)$$

in which *BFL* is an abbreviation for the below-threshold free-running laser. It is interesting to know that we could not calculate the noise fluxes $N_{SP}^{BFL}(\omega)$ and $N_{Pump}^{BFL}(\omega)$ in our previous noise treatment in the absence of an input signal (free-running case) because the fluctuating variable $\delta D(t)$ turned out equal to zero according to (10) of Ref [15].

Although the amplitude noise flux of BFL (27) has a coefficient difference of 4 in numerator with respect to its corresponding quantity (19) of Ref [15], but they have the same bandwidth as

$$(\Delta\omega)_{AM}^{BFL} = 2\gamma_C (1-C),  \qquad (30)$$

which is apparent from the present laser amplifier bandwidth (19) after applying the corresponding conditions $\alpha_S = \beta_S = 0$ and $D_S = CD_0$. The bandwidth (30) was also derived by Loudon and his co-workers (LHSV) by ignoring the cavity Langevin force $\Gamma_\alpha = 0$ and considering the other two Langevin forces of the atomic population inversion $\Gamma_D$ and atomic dipole moment $\Gamma_d$ (please see (4.27) and (4.28) of Ref [13]). By the way, the noise flux conservation of laser amplifier (24) is reduced to

$$4(1-2C)N_{AM}^{BFL}(\omega) + N_{SP}^{BFL}(\omega) = N_{Pump}^{BFL}(\omega),  \qquad (31)$$

so that it can analytically be testified by applying the BFL noise fluxes (27)-(29).

On the other hand, the above-threshold free-running laser (AFL) is distinguished by the different conditions $|\alpha_S|^2 = |\alpha_L|^2 = n_S(C-1)$ and $D_S = D_0$ [8, 13] so that the noise fluxes of amplitude (18), spontaneous emission (21), pumping (26) and their conservation relation (24) exactly reproduce the corresponding AFL noise fluxes (34), (35), (43), and their conservation relation (46) of Ref [15], respectively. The bandwidth of amplitude noise flux (19) is also simplified to

$$(\Delta\omega)_{AM}^{AFL} = 4\gamma_C \frac{C-1}{C},  \qquad (32)$$

which is in complete agreement with the free-running relations (5.62) of LHSV [13] and (47) of Ref [15].

## 5. The DFOTC and the coherence time of class-A laser amplifiers

According to the quantum optics literatures [21, 26], the degree of first-order temporal coherence (DFOTC) of light is defined as a normalized version of the first-order correlation function in the form

$$g^{(1)}(\tau) = g^{(1)}(t'-t) = \frac{\langle \alpha^*(t)\alpha(t')\rangle}{\langle \alpha^*(t)\alpha(t)\rangle} = \frac{\langle \alpha^*(t)\alpha(t+\tau)\rangle}{\langle \alpha^*(t)\alpha(t)\rangle}, \qquad (33)$$

where $\alpha(t)$ is the electric field of an arbitrary optical source whose interference pattern is measured by an optical interferometer with the delay time $\tau = t'-t$ [20]. We have here divided the cavity electric field $\alpha(t)$ into the two parts of statistics amplitude $\alpha_S$ and temporal fluctuating $\delta\alpha_S(t)$ terms according to the trial solution (5). The statics amplitude $\alpha_S$ has already used to study the gain behaviour of class-A and -B laser amplifiers by calculating the normalized mean number of cavity photons $|\alpha_S|^2/n_S$ from the cubic equation (7) [8, 13]. By contrast, our aim here is to study the noise feature and DFOTC of class –A laser amplifiers by using the temporal fluctuating variable $\delta\alpha_S(t)$. In other word, we are looking for a relation between the amplitude noise flux $N_{AM}^{LA}(\omega)$ and the degree of first-order temporal coherence $g_{LA}^{(1)}(\tau)$.

Lets consider the general definition of DFOTC (33) for a laser amplifier as

$$g_{LA}^{(1)}(\tau) = g_{LA}^{(1)}(t-t') = \langle \delta\alpha_S^*(t)\delta\alpha_S(t')\rangle = \frac{1}{2\pi}\int d\omega\, e^{-i\omega t}\int d\omega'\, e^{-i\omega' t}\langle \delta\alpha_S^*(\omega)\delta\alpha_S(\omega')\rangle, \qquad (34)$$

where the neutral role of statics quantity $\alpha_S$ has been cancelled out from numerator and denominator of (33) by inspiring from the trial solution (5). The Fourier integral (34) can be calculated by substituting $\delta\alpha_S(\omega)$ from (13) and implementing the correlation function of cavity Langevin force $\Gamma_\alpha$ given by (3). The result is turned out as

$$g_{LA}^{(1)}(\tau) = \frac{1}{4B}\int_{-\infty}^{\infty} d\omega\, e^{-i\omega\tau}\frac{\gamma/\pi}{\omega^2+\gamma^2} = \frac{1}{4B}\exp(-\gamma|\tau|), \qquad (35)$$

in which $\gamma = \gamma_C B$ is the total damping rate of laser amplifier consists of the mean damping rate of cavity mirrors $\gamma_C$ multiplied by the dimensionless parameter $B$ which appeared as the noise flux bandwidth of laser amplifier in (19). The physical role of DFOTC $g_{LA}^{(1)}(\tau)$ is revealed when it is related to the amplitude noise flux of laser amplifier inside the laser cavity $\left|h_{AM}^{LA}(\omega)\right|^2$ and after emerging from cavity mirrors $N_{AM}^{LA}(\omega)$ by the following Fourier transform (please see (3.5.10) of Ref [26])

$$N_{AM}^{LA}(\omega) = 2\gamma_C \left|h_{AM}^{LA}(\omega)\right|^2 = \frac{2\gamma_C}{\pi} \operatorname{Re}\left[\int_0^\infty d\tau\, e^{i\omega\tau} g_{LA}^{(1)}(\tau)\right]. \tag{36}$$

If one substitutes $g_{LA}^{(1)}(\tau)$ from (35) into (36), a Lorentzian profile is derived for the amplitude noise flux of laser amplifier $N_{AM}^{LA}(\omega)$ in complete agreement with the corresponding relation (18). Meanwhile, the spontaneous emission and pumping noise fluxes are respectively calculated by applying $N_{AM}^{LA}(\omega)$ from (36) into the relations (21) and (26). Therefore, the first advantage of DFOTC $g_{LA}^{(1)}(\tau)$ is to determine the noise fluxes of a class-A laser amplifier.

The other important application of DFOTC is that the coherence time of the emerged light from an optical source $\tau_c$ including the laser amplifier $\tau_c^{LA}$ has a revers relation with the damping rate of DFOTC (35) so that we have (please see (3.4.6) of Ref [26])

$$\tau_c^{LA} = \frac{1}{\gamma} = \frac{1}{\gamma_C B}. \tag{37}$$

A simple comparison of (19) and (37) inspires an uncertainty relation between the bandwidth of the amplitude noise flux $(\Delta\omega)_{AM}^{LA}$ and the coherence time $\tau_c^{LA}$ of laser amplifier as

$$(\Delta\omega)_{AM}^{LA}\, \tau_C^{LA} = 2 = cte. \tag{38}$$

It is reminded that the uncertainty relation (38) is also valid for the bandwidths of spontaneous emission and pumping noise fluxes due to their proportionality with the amplitude noise flux in (21) and (26).

## 6. The DFOTC and the coherence time of free-running class-A lasers

We have so far derived the DFOTC (35) for the light emitted from an amplifying medium with the inverted atomic population inversion such as laser amplifier. The below-threshold free-running laser (BFL) is a special simple case for which the relations (16), (35), (37), and (38) of the class-A laser amplifiers are respectively reduced to

$$B_{BFL} = 1 - C, \tag{39}$$

$$g_{BFL}^{(1)}(\tau) = \frac{1}{4(1-C)} \exp[-\gamma_C(1-C)|\tau|], \tag{40}$$

$$\tau_c^{BFL} = \frac{1}{\gamma_C(1-C)}, \tag{41}$$

and

$$(\Delta\omega)_{AM}^{BFL} = \frac{2}{\tau_C^{BFL}} = 2\gamma_C(1-C), \tag{42}$$

where the corresponding conditions $\alpha_S = \beta_S = 0$, $\omega_d = 0$, and $D_S = CD_0$ have been used.

The confirmation of DFOTC $g_{BFL}^{(1)}(\tau)$ for BFL is straightforward by substituting (40) into the general relation of amplitude noise flux (36) so that a Lorentzian profile will be appeared for the amplitude noise flux of class-A lasers in the below-threshold state and in complete agreement with the corresponding relation (27). In addition, the bandwidth of amplitude noise flux $(\Delta\omega)_{AM}^{BFL} = 2\gamma_C(1-C)$ which has been calculated by applying the coherence time $\tau_c^{BFL}$ from (41) into the uncertainty relation (42) is in consistent with the corresponding relations (4.28) of LHSV [13] and (21) of Ref [15].

The other important case is related to the above-threshold free-running laser (AFL) for which the relations (16), (35), (37), and (38) have respectively simplified to

$$B_{AFL} = \frac{2(C-1)}{C}, \tag{43}$$

$$g_{AFL}^{(1)}(\tau) = \frac{C}{8(C-1)} \exp\left[-\frac{2\gamma_C(C-1)}{C}|\tau|\right], \tag{44}$$

$$\tau_c^{AFL} = \frac{C}{2\gamma_C(C-1)}, \tag{45}$$

and

$$(\Delta\omega)_{AM}^{AFL} = \frac{2}{\tau_C^{AFL}} = \frac{4\gamma_C(C-1)}{C}, \tag{46}$$

where the corresponding conditions $|\alpha_S|^2 = |\alpha_L|^2 = n_S(C-1)$ and $D_S = D_0$ are implemented. Figure 3 illustrates the variations of the normalized bandwidths of amplitude noise fluxes (42) and (46) and their corresponding coherence times (41) and (45) against the normalized pumping rate $C$ in the below-threshold state ($C < 1$) and in the above-threshold state ($C > 1$). The uncertainty relation is apparent throughout the below and above-threshold regions even in their common critical border associated with threshold state ($C = 1$) so that the noise bandwidth tend to zero to compensate the divergent behaviour of the coherence time.

Finally, one can reproduce the above-threshold amplitude noise flux of class-A lasers $N_{AM}^{LA}(\omega)$ given by (18) in a different way by applying $g_{AFL}^{(1)}(\tau)$ from (44) into the amplitude noise flux relation (36) in consistent with (34) of Ref [15]. Similarly, the bandwidth of amplitude noise flux $(\Delta\omega)_{AM}^{AFL} = \frac{4\gamma_C(C-1)}{C}$ which is derived by substituting the coherence time from (45) into the uncertainty relation (46) is in complete agreement with the corresponding relations (5.62) of LHSV [13] and (47) of Ref [15].

## 7. Conclusion

The ability of cavity Langevin force in producing the fluctuations of cavity electric field and atomic population inversion of a single mode class-A laser amplifier is displayed in the both below-threshold state and the injection-locked regime. These fluctuations are rather important because the phase and amplitude noise fluxes are turned out as the correlation function of cavity electric field fluctuation, and the spontaneous emission noise flux is initiated from the atomic population inversion fluctuation. Although the phase noise flux (20) has no special profile but the amplitude and spontaneous emission noise fluxes (18) and (21) demonstrate the Lorentzian profiles with the different heights but the equal bandwidth (19) in agreement with the free-running case (47) of Ref [15]. It is also revealed that the noise bandwidth of laser amplifier (19) are varied by the input signal parameters of amplitude and frequency detuning as well as the laser parameters of pumping and cavity mirrors damping rates. The main source of output amplitude and spontaneous emission noise fluxes and is due to the input pumping noise flux which are related together by the flux conservation relation (24).

The degree of first-order temporal coherence (DFOTC) is defined by (34) as the correlation function of the cavity electric field fluctuation in the time domain. The first physical importance of DFOTC is that its damping rate has the reverse relation (37) with respect to the coherence time of light emitted from the laser amplifier. The second one is concerned to its Fourier transform which presents the amplitude noise flux according to the relation (36). Finally, the coherence time (37) and the bandwidth of amplitude noise flux (19) are dependent quantities due to the uncertainty relation (38). By increasing the input signal detuning or reducing the input signal flux, the bandwidth of amplitude noise flux is shorten as illustrated in Figs. 1 and 2.

The novel results of class-A laser amplifiers including the amplitude, spontaneous emission, and pumping noise fluxes, their conservation relation, the DFOTC, the coherence time, and the latter uncertainty with the bandwidth of amplitude noise flux are directly converted to those of free-running class-A lasers by ignoring from the amplitude ($\beta_S = 0$) and frequency detuning ($\omega_d = 0$) of the input signal. In this way, the noise fluxes (27)-(29) and the optical characteristics (39)-(42) of BFL are derived from the corresponding below-threshold relations of laser amplifier after applying the conditions $\alpha_S = 0$ and $D_S = CD_0$ which are valid for the normalized pumping rate $C < 1$. Similarly, the noise fluxes of AFL together with the optical characteristics (43)-(46) are derived from the injection-locked relations of laser amplifier after applying the conditions $|\alpha_S|^2 = |\alpha_L|^2 = n_S(C-1)$ and $D_S = D_0$ which are valid for the normalized pumping rate $C > 1$.

In the end, the noise feature of multiple-frequency regime is under progress and the results will be announced in future soon.

**Figure Captions**

**Fig. 1.** The Lorentzian profiles of amplitude noise flux are illustrated for the different normalized frequency detuning $\omega_d/\gamma_C$ shown inside the figure, the typical normalized input flux $\gamma_2|\beta_S|^2/n_s = 0.25$, and normalized pumping rates (a)- $C = 0.5$ associated with the below-threshold state and (b)- $C = 1.2$ associated with the above-threshold (injection-locked) state.

**Fig. 2.** The opposite role of normalized input signal flux and frequency detuning on the bandwidth of amplitude noise flux is demonstrated for the normalized pumping rates (a)- $C = 0.5$ and (b)- $C = 1.2$. The noise bandwidth of laser amplifier is evidently shorten by increasing the normalized frequency detuning $\omega_d/\gamma_C$ which is in agreement with Fig. 1. The noise bandwidths of free-running laser have also displayed on the vertical axes by the single points 1 and 0.67 which are calculated by using the relations (30) and (32) corresponding to the below and above threshold states, respectively.

**Fig. 3.** The simultaneous variations of noise bandwidth (red colour) and coherence time (blue colour) versus the normalized pumping rate $C$ are plotted for the free-running laser in the below and above threshold states. The uncertainty principle between the noise bandwidth and coherence time is evident in all the regions even at the threshold state where the noise bandwidth tends to zero to compensate the divergent behaviour of coherence time.

Fig. 1 (a)

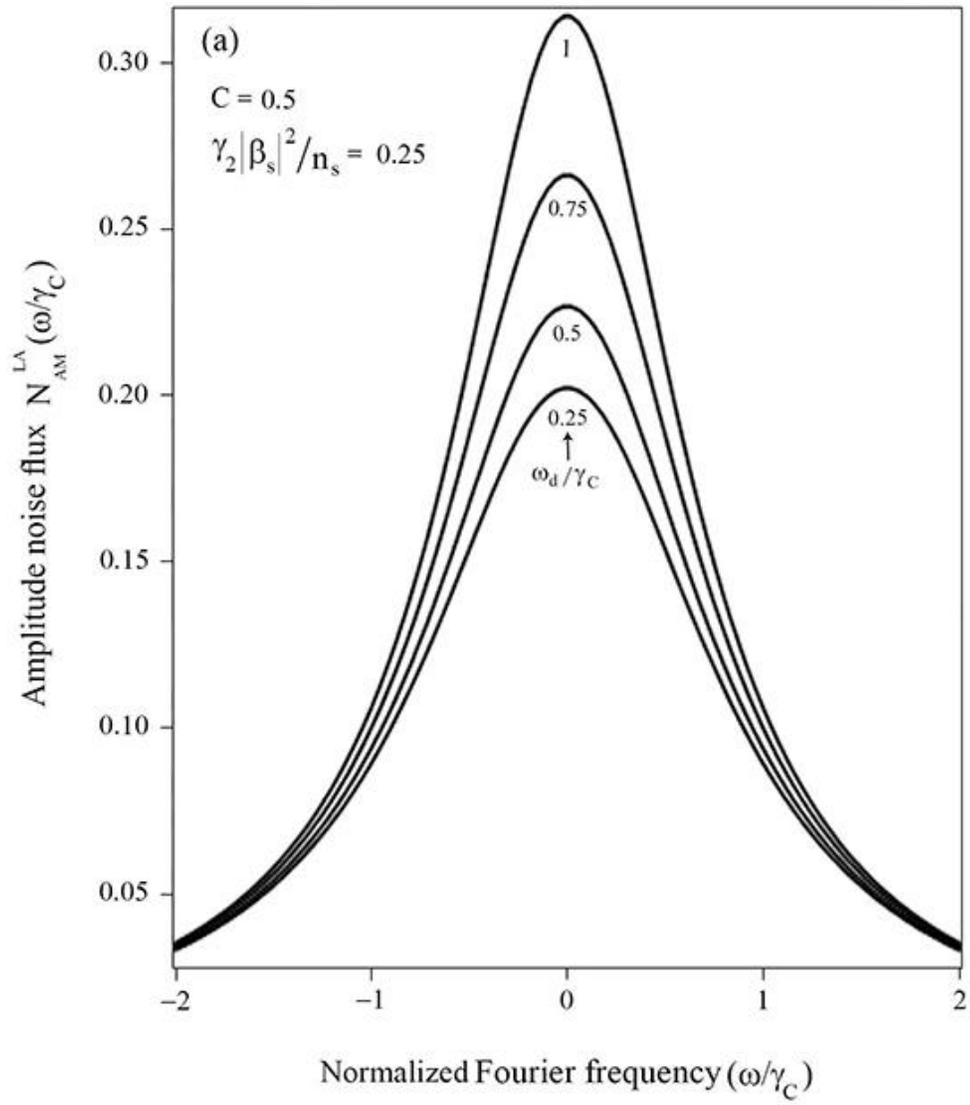

Fig. 1 (b)

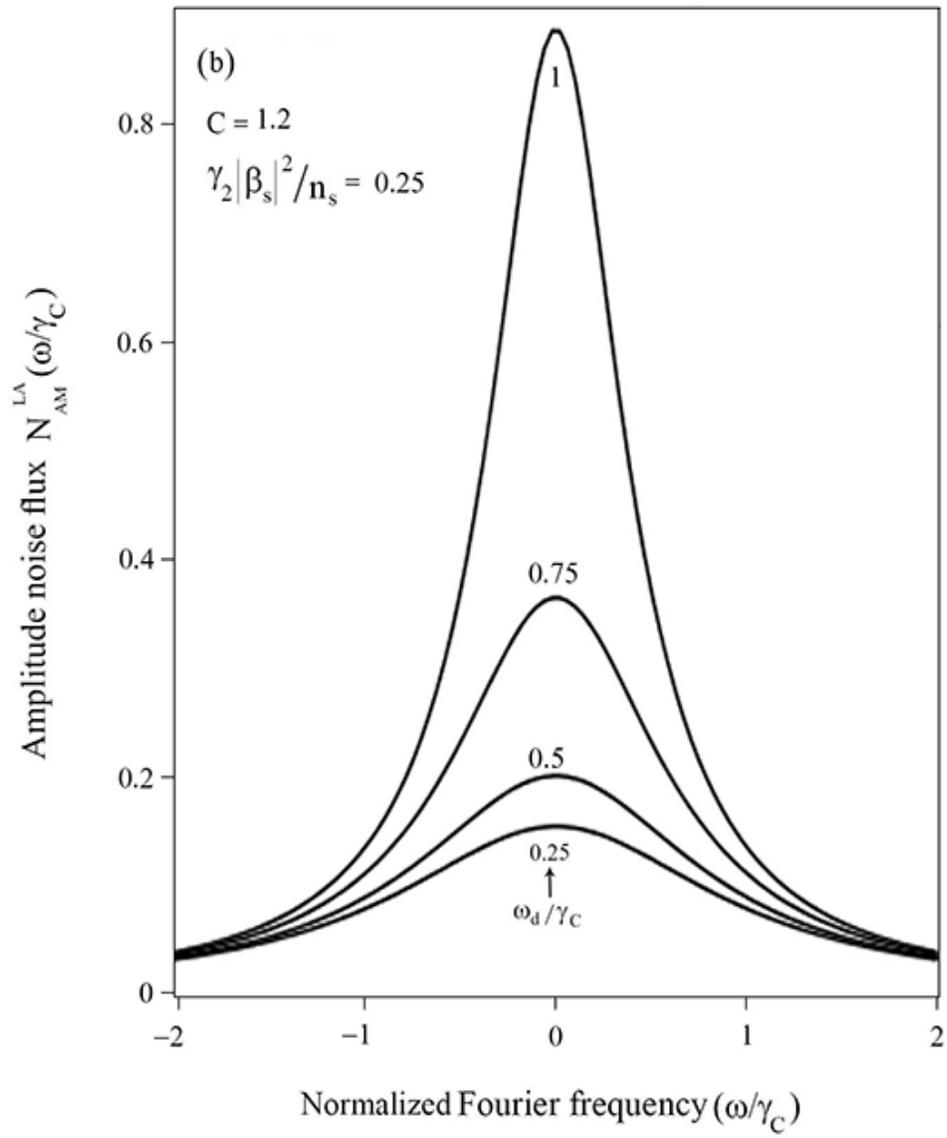

Fig. 2 (a)

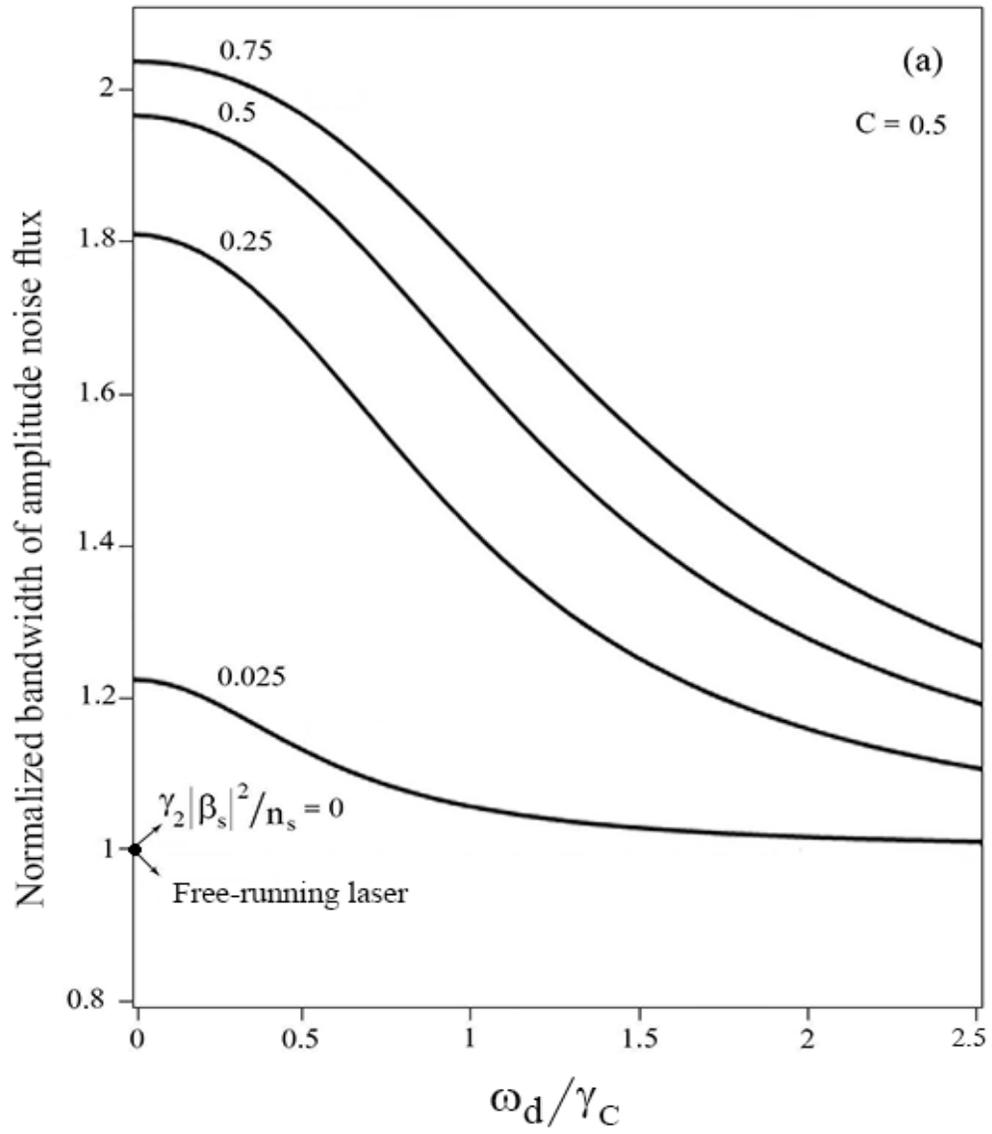

Fig. 2(b)

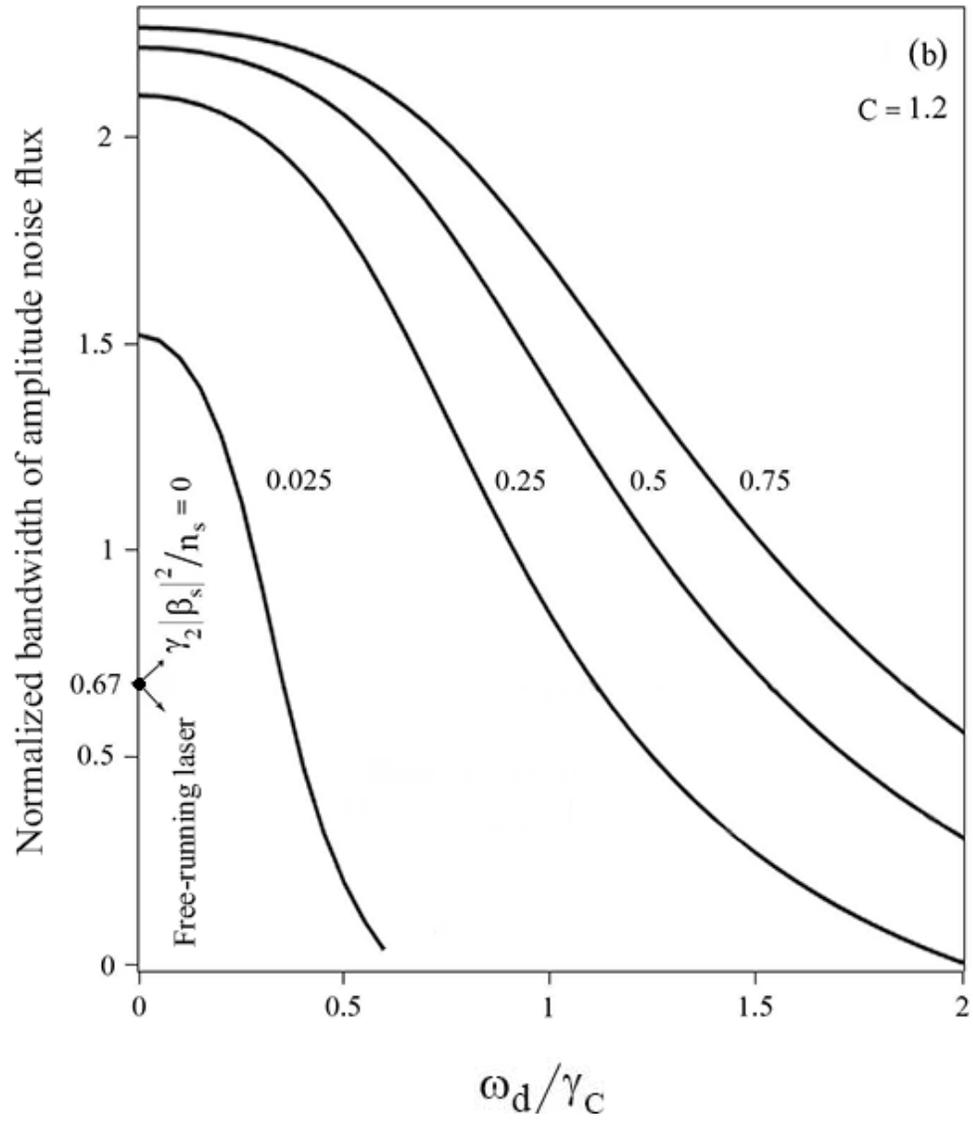

Fig. 3

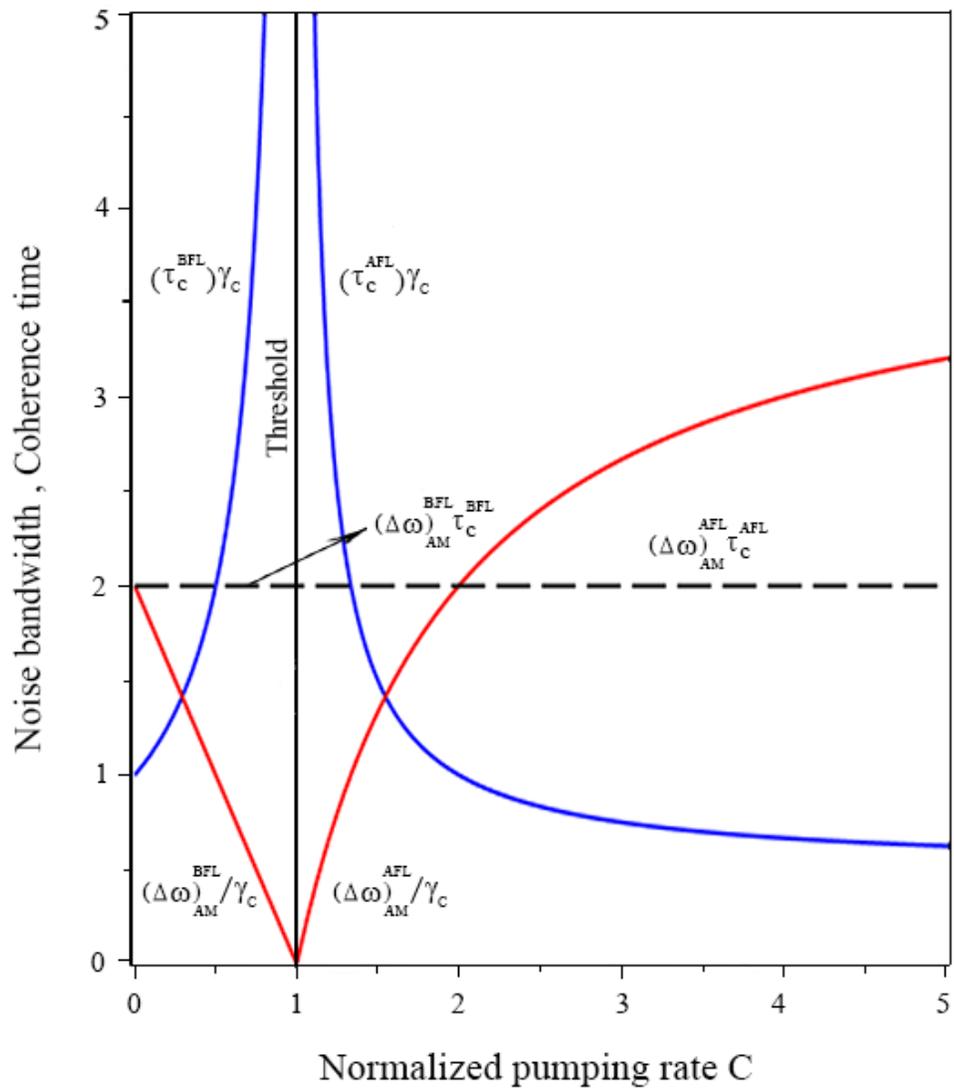